\documentclass[english, format=sigchi, anonymous=false, balance=false]{acmart}

\renewcommand\footnotetextcopyrightpermission[1]{} 
\usepackage[T1]{fontenc}
\usepackage[latin9]{inputenc}
\usepackage{array}
\usepackage{natbib}
\usepackage{booktabs}
\usepackage{multirow}
\usepackage{amssymb}
\usepackage{graphicx}
\usepackage{babel}

\makeatletter

\providecommand{\tabularnewline}{\\}

\usepackage[font=small,skip=2pt]{caption}

\makeatother

\begin{document}
\global\long\def\cov#1#2{\mathrm{cov}(#1,#2)}
\fancyhead[L, R]{}

\title{On the Effectiveness of Low-rank Approximations for Collaborative
Filtering compared to Neural Networks}

\author{Marcel Kurovski}
\affiliation{inovex GmbH}
\email{mkurovski@inovex.de}
\authornote{Both authors contributed equally to this work}

\author{Florian Wilhelm}
\authornotemark[1]
\affiliation{inovex GmbH}
\email{fwilhelm@inovex.de}

\begin{abstract}
Even in times of deep learning, low-rank approximations by factorizing
a matrix into user and item latent factors continue to be a method
of choice for collaborative filtering tasks due to their great performance.
While deep learning based approaches excel in hybrid recommender tasks
where additional features for items, users or even context are available,
their flexibility seems to rather impair the performance compared
to low-rank approximations for pure collaborative filtering tasks
where no additional features are used. Recent works propose hybrid
models combining low-rank approximations and traditional deep neural
architectures with promising results but fail to explain why neural
networks alone are unsuitable for this task. In this work, we revisit
the model and intuition behind low-rank approximation to point out
its suitability for collaborative filtering tasks. In several experiments
we compare the performance and behavior of models based on a deep
neural network and low-rank approximation to examine the reasons for
the low effectiveness of traditional deep neural networks. We conclude
that the universal approximation capabilities of traditional deep
neural networks severely impair the determination of suitable latent
vectors, leading to a worse performance compared to low-rank approximations.
\end{abstract}

\keywords{Recommender Systems; Neural Networks; Collaborative Filtering}
\ccsdesc[500]{Information systems~Collaborative filtering}
\ccsdesc[500]{Computing methodologies~Neural networks}
\ccsdesc[500]{Computing methodologies~Factorization methods}

\maketitle

\section{Introduction}

Since the Netflix prize in 2009, variants of low-rank approximation
(LRA) have been and still are among the most popular approaches to
collaborative filtering (CF) problems despite the advent of Deep Learning
(DL). In many other domains, e.g. computer vision, image and speech
recognition, classical methods used in those domains were significantly
surpassed by neural networks and thus great progress was made. Although
neural network based recommender systems have become widespread and
shown significant performance gains by exploiting content, contextual
and sequential patterns, they prove insufficient in case of CF. Feature
engineering and extraction capabilities of DL seem to be of no use
and the sparsity of user-item interactions surely is an impeding factor.
This is generally acknowledged by the community and neural networks
are often combined with LRA approaches\citep{He2017c,Guo2013}. Despite
these works, we found no proper study comparing LRAs with neural networks
especially with respect to the determined latent vectors by these
methods. Our work remedies this by contributing an intuition for LRAs
based on expected covariances between latent features and interactions.
Moreover, we evaluate in several experiments the suitability of the
latent vectors obtained by low-rank approximations compared to neural
networks.

\subsection{Problem Formulation}

Given users $u_{i}$ with $i=1,\ldots,m$ and items $v_{j}$ with
$j=1,\ldots,n$ from a single domain, we denote with a scalar$r_{ij}$
the interaction of user $u_{i}$ with item $v_{j}$. In order to express
the user's preference for an item, we assume $r_{ij}<r_{ik}$ if $u_{i}$
liked $v_{k}$ more than $v_{i}$ and $r_{ij}=r_{ik}$ if $u_{i}$
is indifferent between $v_{i}$ and $v_{k}$. In implicit feedback
scenarios, we often have $r_{ij}=1$ for positive feedback, whereas
in explicit feedback scenarios $r_{ij}$ takes a numerical rating.
We denote the interaction matrix of all users and items with $R=\{r_{ij}\}$.

\subsection{Low-Rank Approximations\label{subsec:low_rank_approx}}

LRAs exploit the fact that rows and columns of $R$ are highly correlated
due to redundancies in the underlying ratings, e.g. similarly acting
users or similarly rated items. This allows a robust approximation
by a lower-rank matrix $\hat{R}=\{\hat{r}_{ij}\}$ \citep{Ricci2015,Aggarwal2016}.
We have

\[
\begin{array}{c}
\hat{r}_{ij}:=\mathbf{e}_{i}^{u}\cdot\mathbf{e}_{j}^{v}+b_{i}^{u}+b_{j}^{v},\mathbf{e}_{i}^{u},\mathbf{e}_{j}^{v}\in\mathbb{R}^{p},p\in\mathbb{N},b_{i}^{u},b_{j}^{v}\in\mathbb{R},\end{array}
\]

where $\mathbf{e}_{i}^{u}$ and $\mathbf{e}_{j}^{v}$ are elements
of a joint latent space of dimension $p\ll\min\{m,n\}$ denoted as
user or item\textit{ latent factors}\textit{\emph{ in a LRA context}}
or more generally\emph{ latent vectors}. Their inner product models
the user-item interaction. The bias terms $b_{i}^{u}$ and $b_{j}^{v}$
capture interaction-independent effects like users systematically
rating lower than others or items that are more popular than others.
The actual approximation depends on the problem setting. For explicit
feedback, where $r_{ij}\in\{-1,1\}$ or $r_{ij}\in\{0,1\}$, a pointwise
approach is often applied and for instance the binary cross-entropy
loss, i.e.
\begin{equation}
-\sum_{S}\left[r_{ij}'\cdot\log(\sigma(\hat{r}_{ij}))+(1-r_{ij}')\cdot\log(1-\sigma(\hat{r}_{ij}))\right],\label{eq:cross_entropy}
\end{equation}
where $r_{ij}'=\max\{0,r_{ij}\}$, $\sigma$ is the sigmoid function
and $S:=\{(i,j\}|r_{ij}\:\mathrm{is\:known}\}$, is optimized. In
case of implicit positive feedback, the de facto standard is Bayesian
Personalized Ranking (BPR). This pairwise approach maximizes the probability
that an item $v_{j}$ with observed interaction $r_{ij}$ of user
$i$ is ranked higher than an item$v_{k}$ with no observed interaction
$r_{ik}$ with an item $k$, i.e. $p(v_{j}>_{u_{i}}v_{k}\mid\mathbf{e}_{i}^{u}\mathbf{e}_{j}^{v},\mathbf{e}_{k}^{v},b_{i}^{u},b_{j}^{v})$
\citep{Rendle2009}. The model parameters are determined with the
help of (stochastic) gradient descent or alternating least squares
methods \citep{Hu2008,Koren2008b}.

\subsection{Neural Network Approaches}

In recent years, DL based recommenders have become widespread in academia
and industry \citep{Zhang2018}. Leveraging flexible, non-linear models
for representation learning and sequence modeling has proven highly
beneficial, especially in content-based and hybrid settings. However,
some works also use neural networks, e.g. multi-layer perceptrons
(MLP), autoencoders, and convolutional neural networks, purely for
CF \citep{HaoWangNaiyanWang2014,He2017c,He2018a,Song2018a}. This
invites to research their effectiveness compared to LRAs.

A general MLP-based model, illustrating the basic structure of neural
collaborative filtering networks (NCFN), is

\[
f_{\Theta,U,V}(u_{i},v_{j})=g(h_{u}(u_{i}),h_{v}(v_{j})\mid U,V)\mid\Theta),
\]

where $f$ is a mapping from a user-item tuple $(u_{i},v_{j})$ into
$\mathbb{R}$. It is composed of functions $h_{u},h_{v}$ that transforms
user and item indices into their joint latent space and a MLP $g$
with parameters $\Theta$ which maps the concatenation, outer product
or Hadamard product of these latent vectors into $\mathbb{R}$ modeling
the user-item interaction \citep{He2017c,He2018a,Song2018a}. Since
neural networks are universal function approximators, they are theoretically
capable to derive these interactions when the concatenation of user
and item latent vectors are provided as inputs. Analogously to LRAs,
NCFNs can be trained in an explicit context using the binary cross-entropy
(\ref{eq:cross_entropy}) or in an implicit context with BPR. The
parameters are inferred by using backpropagation for loss minimization
and hence weight adaption by means of (stochastic) gradient descent.

\section{Covariances and low-rank approximations\label{sec:Covariances-and-low-rank}}

We want to establish the connection between LRAs and the covariances
of latent vectors and interactions to give a novel intuition behind
the LRA model. In case of a purely CF task, $u_{i}$ as well as $v_{j}$
are just entities without any observable features. Thus we assume
the existence of latent item features $l_{k}$ with $k=1,\ldots,p$
which describe the items since we assume a single domain. Consequently,
we have for each item $v_{j}$ a latent vector $\mathbf{e}_{j}^{v}\in\mathbb{R}^{p}$
where $e_{jk}^{v}$ defines how strong the latent feature $l_{k}$
is prevalent in $v_{j}$. Analogously, we have for each user $u_{i}$
a latent vector $\mathbf{e}_{i}^{u}\in\mathbb{R}^{p}$ where $e_{ik}^{u}$
represents the strength of the preference for $l_{k}$.

Under these assumptions, we can conclude that if user $u_{i}$ has
a preference for $l_{k}$ this will also be reflected in the items
the user has interacted with. With $R_{i}^{u}$ we denote the random
variable for the interactions of user $u_{i}$ having realizations
$r_{ij}$ and analogously the random variable for the prevalence of
the feature $k$ in items with $E_{k}^{v}$ and realizations $e_{jk}^{v}$.
We can now formalize the joint variability of $R_{i}^{u}$ and the
prevalence of $l_{k}$ in the items, $u_{i}$ has interacted with,
as $\cov{R_{i}^{u}}{E_{k}^{v}}>0$. Following the same reasoning but
from an item's perspective, we can argue that the preference of users
for $l_{k}$ should be reflected in their interactions with an item
$v_{j}$ having a strong prevalence of $l_{k}$ and thus we have that
$\cov{R_{j}^{v}}{E_{k}^{u}}>0$. If we now interpret $r_{ij}$ as
realizations of a random variable $R$, we can express these relationships
jointly as $\cov R{E_{k}^{u}E_{k}^{v}}>0$ due to the fact that $E_{k}^{u}$
and $E_{k}^{v}$ can be assumed independent.

Using the bilinearity of the covariance, we have for all latent item
features that
\begin{equation}
\cov R{\sum_{k=1}^{p}\alpha_{k}E_{k}^{u}E_{k}^{v}}>0,\label{eq:pos_cov}
\end{equation}
where $\alpha_{k}>0$ weights the importance of $l_{k}$ with respect
to the other latent features. Having derived this canonical condition
allows us to justify many traditional methods for  CF tasks. For instance
classical matrix factorization based methods in an implicit feedback
use-case, fulfill (\ref{eq:pos_cov}) by assuming equal importance
of $l_{k},$ i.e. setting $\alpha_{k}=1$ for $k=1,\ldots,p$ and
determining $e_{ik}^{u}$ and $e_{jk}^{v}$ such that $\hat{r}_{ij}=\sum_{k=1}^{p}e_{ik}^{u}e_{jk}^{v}=\mathbf{e}_{i}^{u}\cdot\mathbf{e}_{j}^{v}$.

\section{Experiments}

In the following section we provide empirical validation of the covariance
intuition together with an extensive comparison between LRA and DL.\footnote{The source code is available at https://github.com/FlorianWilhelm/lrann.}

\subsection{Dataset}

We use the MovieLens 100k dataset \citep{Harper2015a} for our empirical
study. The dataset contains $100,836$ ratings between $m=610$ users
and $n=9,724$ items on a discrete rating scale with $r_{ij}\in\{0.5,1.0,...,5.0\}$.
We provide an implicit and explicit interpretation of the rating data
to analyze the results in both feedback scenarios.

For the\emph{ implicit feedback} scenario, we only keep all interactions
rated equal and above each user's mean rating labeled with $1$ and
set all others to $0$. This yields a remainder of $54,732$ ratings.
We use $BPR$ loss for training which also maximizes the AUC \citep{Dhanjal2015}.
$BPR$ randomly samples negative feedback from the remaining unobserved
items for each user. In the\emph{ explicit feedback} scenario, we
binarize the original ratings using the users' mean ratings as threshold
resulting in labels $1$ and $-1$. We then use the binary cross-entropy
(\ref{eq:cross_entropy}) to fit the data to our models.

\subsection{Methodology}

\subsubsection{Covariances and LRAs}

In both scenarios, we fit latent vectors of size $p=32$ with minibatch
gradient descent (batch size 128) for 15 training epochs using the
Adam Optimizer \citep{Kingma2014}. We use a learning rate $\alpha=0.003$,
exponential decay rates $\beta_{1}=0.9$ and $\beta_{2}=0.999$ and
no regularization. Hence, we obtain latent vectors $\mathbf{e}_{i}^{u}$
and $\mathbf{e}_{j}^{v}$. We calculate the covariance $\cov{R_{i}^{v}}{E_{k}^{u}}$
between the interactions $r_{ij}$ of a fixed user $i$ and latent
vectors component $e_{jk}^{v}$ of the items she interacted or not
interacted with. The covariances over varying $k$ are then correlated
to the user's latent vector using Pearson which we denote as $\rho_{i}^{u}:=\rho(E_{k}^{u},\cov{R_{i}^{u}}{E_{k}^{v}})$.
Analogously, we define the correlation $\rho_{j}^{v}:=\rho(E_{k}^{v},\cov{R_{j}^{v}}{E_{k}^{u}})$
from the view of a fixed item and varying users that interacted or
not interacted with it. Highly positive correlation provides empirical
support for our theory from Section (\ref{sec:Covariances-and-low-rank}).

\subsubsection{NCFN}

In order to compare LRA and DL, we limit experiments to implicit feedback
due to its greater abundance in real-world applications and competitive
results by using the pairwise ranking approach BPR. Essentially, we
examine to which extent network input modeling as well as pretraining
latent vectors influence the capability of a DNN to reproduce or potentially
outperform a strong LDA baseline in terms of accuracy as measured
by the Mean Reciprocal Rank (MRR), Mean Average Precision at 10 (MAP@10)
and Area Under receiver operating characteristic Curve (AUC). We explore
different DNN architectures and distinguish between three pretraining
strategies. $DNN$ refers to the first setting where the latent vectors
are initialized randomly and constitute the parameter space to fit
together with the network parameters. We augment this setting by initializing
the latent vectors to those of our LDA baseline model. In this case
we distinguish whether the latent vectors can still be adjusted or
whether they stay fixed, yielding $DNN_{pretrained}$ and $DNN_{pretrained}^{fixed}$.
Input modeling wise, we separate between feeding concatenated user-item
latent vectors $[\mathbf{e}_{i}^{u},\mathbf{e}_{j}^{v}]$ and their
Hadamard product $\mathbf{e}_{i}^{u}\odot\mathbf{e}_{j}^{v}$, similar
to \citep{He2017c,He2018a}, into the network.

For each of the resulting six combinations we explore neural networks
with $L\in\{0,1,2,3\}$ hidden layers, using different activation
functions $\{ReLU,ELU,\tanh,sigmoid\}$ arriving at 13 combinations.
We choose $\alpha\in\{{0.001,0.003,0.01}\}$ and five different random
initializations for the network parameter initialization. Each combination
is trained for 20 epochs. We compute test set MRR after every epoch
for early stopping.\footnote{Thus, with $2\times3$ settings, $13$ neural network architectures,
$3\times5\times20$ training epochs, we consider $23,400$ experiments.} For sake of brevity of our study, we leave other goals of recommendations
as diversity, serendipity for future work. We apply a 80/20 train-test
split and keep it consistent across LRA and NCFN.\footnote{We perform a hyperparameter grid search to find the best and therefore
most competitive LRA configuration using MRR as selection criterion.}

\begin{figure}[t]
\begin{centering}
\includegraphics[width=\columnwidth]{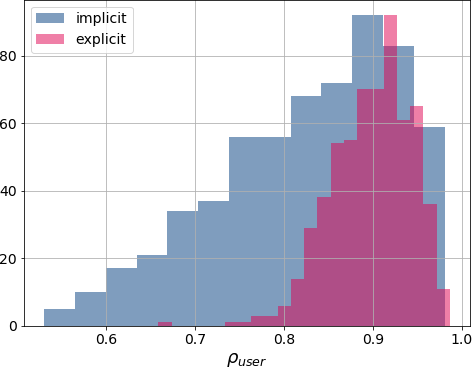}
\par\end{centering}
\caption{Histogram for user correlations in an\emph{ implicit} and\emph{ explicit}
scenario $\rho(E_{k}^{u},\protect\cov{R_{i}^{u}}{E_{k}^{v}})$. \label{fig:user_corr_joint}}
\bigskip
\begin{centering}
\includegraphics[width=\columnwidth]{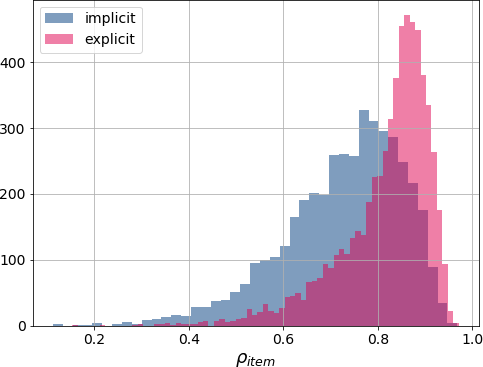}
\par\end{centering}
\caption{Histogram for item correlations in an\emph{ implicit} and\emph{ explicit}
scenario $\rho(E_{k}^{v},\protect\cov{R_{j}^{v}}{E_{k}^{u}})$. \label{fig:item_corr_joint}}
\end{figure}

\subsection{Results and Discussion}

\subsubsection{Covariances and LRAs}

Evaluating $\rho_{i}^{u},i\in I$ as well as $\rho_{j}^{v},j\in J$
for both feedback scenarios, we observe highly positive correlations
with $\rho_{mean}^{u}=0.8246$ and $\rho_{mean}^{v}=0.7256$ (implicit
feedback) and $\rho_{mean}^{u}=0.8987$ and $\rho_{mean}^{v}=0.8156$
(explicit feedback) as detailed in Table \ref{tab:corr_stats}. Due
to the fact that we only consider significant individual correlations
and that many items have just few or no interactions, there remains
just a fraction of items in each scenario. The distributions of these
correlations are also shown in Figures \ref{fig:user_corr_joint}
and \ref{fig:item_corr_joint}. These results empirically support
our derivation of LRA based on covariances of users' preferences and
items' features manifested by user-item interactions.

\begin{table}
\begin{centering}
\begin{tabular}{ccccc}
\toprule 
 & \multicolumn{2}{c}{Implicit} & \multicolumn{2}{c}{Explicit}\tabularnewline
\midrule
\midrule 
 & user & item & user & item\tabularnewline
\midrule 
$n$ & 610 & 4275 & 610 & 6278\tabularnewline
\midrule 
$mean$ & \textbf{0.8246} & \textbf{0.7256} & \textbf{0.8987} & \textbf{0.8156}\tabularnewline
\midrule 
$\sigma_{\rho}$ & 0.1011 & 0.1275 & 0.0446 & 0.098\tabularnewline
\midrule 
\textbf{$\rho_{min}$} & 0.5304 & 0.112 & 0.6589 & 0.1534\tabularnewline
$\rho_{0.25}$ & 0.7552 & 0.6505 & 0.8683 & 0.7752\tabularnewline
$\rho_{0.5}$ & \textbf{0.8425} & \textbf{0.7472} & \textbf{0.9034} & \textbf{0.8433}\tabularnewline
$\rho_{0.75}$ & 0.9061 & 0.8207 & 0.9318 & 0.8827\tabularnewline
\textbf{$\rho_{max}$} & 0.9806 & 0.9681 & 0.9867 & 0.9710\tabularnewline
\bottomrule
\end{tabular}
\par\end{centering}
\centering{}\medskip{}
\caption{Correlation statistics for user and item views in an \emph{implicit}
and \emph{explicit} feedback scenario. \label{tab:corr_stats}}
\end{table}

\subsubsection{NCFN}

\begin{figure}
\begin{centering}
\includegraphics[width=\columnwidth]{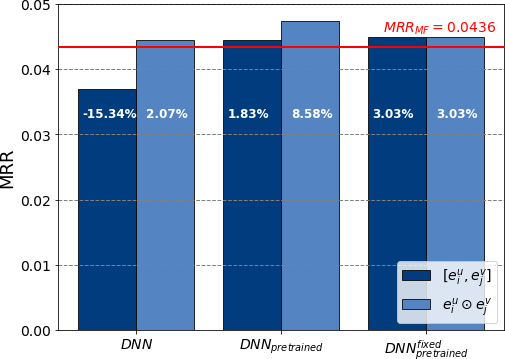}
\par\end{centering}
\caption{Test set MRR for our best models by input (concatenation or Hadamard
product) and pretraining strategies compared to the best MRR obtained
from hyperparameter-optimized LRA.\label{fig:nn_search_results}}

\end{figure}

A traditional DNN with concatenation of user and item latent vectors
could not match the performance of LRA in our experiments. This result
is in line with \citet{Dziugaite2015} who stated ``Conceivably,
a deep neural network could learn to approximate the element-wise
product or even outperform it, but this was not the case in our experiments,
which used gradient-descent techniques to learn the neural network
weights.`` Despite the universal approximation capability of such
a network, we need to explicitly model the inner product of latent
user and item vectors to match LRA's performance. We believe that
the flexibility of a traditional DNN impedes the determination of
proper latent vectors. This hypothesis is supported by our experiments
that use pretrained latent vectors from LRA. In this case even a DNN
with concatenation is on par with LRA, and even more so, if the pretrained
latent vectors are fixed. The largest performance gain is achieved,
however, by explicitly modeling the user-item interaction with the
help of the Hadamard product resulting in a boost of 20\% in MRR.
Combining the fitted latent factors of the best LRA that already embed
the underlying preference relations with the Hadamard product even
outperforms LRA to a certain extent, i.e. 8.58\% in $DNN_{pretrained}$.
This can be interpreted as an adaption of $\alpha_{k}$ in (\ref{eq:pos_cov})
by the neural network which is more flexible as setting $\alpha_{k}=1$.
All results are summarized in Figure \ref{fig:nn_search_results}
and Table \ref{tab:nn_search_results}.

\begin{table}
\begin{centering}
\begin{tabular}{clccc}
\cmidrule{3-5} 
 &  & \textbf{MRR} & \textbf{MAP@10} & \textbf{AUC}\tabularnewline
\midrule 
\multicolumn{2}{c}{\textbf{$MF_{best}$}} & 0.0436 & 0.0706 & 0.9211\tabularnewline
\midrule
\midrule 
\multirow{3}{*}{$[\mathbf{e}_{i}^{u},\mathbf{e}_{j}^{v}]$} & $DNN$ & 0.0369 & 0.0627 & 0.8920\tabularnewline
 & $DNN_{pretrained}$ & 0.0444 & \textbf{0.0738} & \textbf{0.9195}\tabularnewline
 & $DNN_{pretrained}^{^{fixed}}$ & \textbf{0.0449} & 0.0706 & 0.9138\tabularnewline
\midrule
\midrule 
\multirow{3}{*}{$\mathbf{e}_{i}^{u}\odot\mathbf{e}_{j}^{v}$} & $DNN$ & 0.0445 & 0.0716 & 0.9157\tabularnewline
 & $DNN_{pretrained}$ & \textbf{0.0473} & \textbf{0.0753} & \textbf{0.9241}\tabularnewline
 & $DNN_{pretrained}^{^{fixed}}$ & 0.0449 & 0.0713 & 0.9216\tabularnewline
\bottomrule
\end{tabular}
\par\end{centering}
\caption{Comparison between our best DNN and LRA models with respect to different
strategies for user-item latent vectors, e.g. $pretrained$ and/or
$fixed$, and concatenation $[\cdot,\cdot]$, resp. Hadamard product
$\odot$, in terms of MRR, MAP@10 and AUC.\label{tab:nn_search_results}}

\end{table}

\section{Conclusion}

This work contributes theoretical and empirical studies examining
the effectiveness of LRA compared to DNNs. We showed that standard
DNNs fail to approximate element-wise multiplications which is the
cornerstone of LRA's effectiveness according to our model derivation
using covariances. Traditional DNNs perform significantly worse than
LRAs for CF. However, when using proper initialization of the latent
vectors from a pretrained LRA and potentially joining them using the
Hadamard product, DNNs can outperform LRAs. These are important insights
to consider when designing DNN based recommender systems that (partially)
depend on collaborative signals. Our results are also supported by
latest works that show surprising incapacities of neural networks.
For example, \citet{Trask2018} propose a neural arithmetic logic
unit (NALU) to alleviate the fact that DNNs fail to systematically
abstract and to extrapolate from the provided training data. \citet{Lin2017a}
propose dedicated multiplication gates to enable DNNs solving a seemingly
simple task.

For future work, we want to further deepen the understanding of DNNs
for CF tasks compared to LRAs. We believe that the concurrent adaption
of both, latent vector space and neural network parameters, leads
to suboptimal configurations which is supported by the gained performance
when we used pretrained latent vectors in our experiments. Eventually,
a deeper understanding of these inner workings will result in more
advanced DNN-based recommenders.\pagebreak{}

\bibliographystyle{ACM-Reference-Format}
\bibliography{references}

\end{document}